\documentclass[10pt,prb,aps,twocolumn,superscriptaddress,showpacs,amsfonts,amsmath,amssymb,showkeys,floatfix]{revtex4-1}

\usepackage{bm}
\usepackage{pstricks}
\usepackage{setspace}
\usepackage{fdsymbol}
\usepackage{amssymb}
\usepackage{graphicx}

\begin{document}

\title{Role of positional disorder in fully textured ensembles of Ising-like dipoles}

\date{\today}
\author{Juan J. Alonso}
\email[e-mail address: ] {jjalonso@uma.es}
\affiliation{F\'{\i}sica Aplicada I and Instituto Carlos I  de F\'{\i}sica Te\'orica y Computacional, Universidad de M\'alaga, 29071 M\'alaga, Spain}
\author{B. All\'es}
\email[E-mail address: ] {alles@pi.infn.it}
\affiliation{INFN--Sezione di Pisa, Largo Pontecorvo 3, 56127 Pisa, Italy}
\author{J.G. Malherbe}
\email[E-mail address: ] {malherbe@u-pec.fr}
\author{V. Russier}
\email[E-mail address: ] {vincent.russier@cnrs.fr}
\affiliation{ICMPE, UMR 7182 CNRS and UPE 2-8 rue Henri Dunant 94320 Thiais, France.}

\date{\today}

\begin{abstract}
 We study by numerical simulation the magnetic order in ensembles of randomly packed  magnetic spherical particles which, induced by their uniaxial anisotropy in the strong coupling limit,
behave as Ising dipoles.  We explore the role of the frozen disorder in the positions of the particles assuming a common fixed direction for the easy axes of all spheres.
We look at  two types of  spatially disordered configurations.  In the first one we consider isotropic positional distributions which can be obtained from the liquid
state of the hard sphere fluid. We derive the phase diagram  in the $T$-$\Phi$ plane where $T$ is the temperature and $\Phi$ the volume fraction.
This diagram exhibits long-range ferromagnetic order at low $T$ for volume fractions above the threshold $\Phi_{c} = 0.157$ predicted by mean-field calculations. 
For $\Phi \lesssim \Phi_{c}$ a spin-glass phase forms with the same marginal behavior found for other strongly disordered dipolar systems.
The second type of   spatial configurations we study are anisotropic 
distributions that can be obtained by freezing a dipolar hard sphere liquid in its polarized state at low temperature.
This  structural anisotropy enhances the ferromagnetic order present in isotropic hard sphere configurations.

\end{abstract} 
\maketitle

\section{INTRODUCTION}
\label{intro}
Advances in nanotechnology have permitted to synthesize nanoparticles (NP) of various sizes and shapes, 
with or without non-magnetic coating layers,
and  to create monodisperse systems of NP with a certain control on their spatial distribution.\cite{nano}
Small enough NP (with diameters of about 10-30 nm) have a single domain that behaves like a magnetic dipole.\cite{skomski} 

Furthermore the internal structure of the NP gives rise to the magnetocrystalline anisotropy energy (MAE)
tending to orient the dipoles along local easy axes.
Under such a circumstance magnetic spin flips have  a non-vanishing energetic cost $E_{a}$.\cite{bedanta, fiorani}
For sufficiently dense packings, the interaction energy $E_{dd}$ between nearby dipoles can be comparable to $E_{a}$.
For example for  compact packings of bare maghemite nanoparticles the ratio $E_{a}/E_{dd}$ is approximately\cite{small} $E_{a}/E_{dd} \approx 6$. 
For such systems $E_{dd}$ can induce complex collective behavior endowing them with a rich phenomenology, mainly at low temperature.\cite{fiorani,toro1,sawako}.  
Indeed, the spatial variations of the dipolar fields  lead to geometric frustration, 
making these systems rather sensitive to the relative positions and directions of their dipoles. For example, dipoles placed in  well-ordered
crystalline arrangements  exhibit ferro- (FM) or antiferromagnetic (AF) order depending on the lattice geometry. \cite{luttinger} 

In the strong MAE limit 
the dipole of each NP  points up or down nearly parallel to its local easy axis 
leading to a dipolar Ising-like model \cite{knak}  
where only $E_{dd}$ play a central role.

The study of the magnetic order in systems of magnetic NP is an active field of research.\cite{bookpeddis, batlle}
Its interest ranges from fundamental physics, because of the need of understanding the collective effects of samples of NP, to their applications
into a broad class of technological problems like data storage\cite{dobson} or nanomedicine.\cite{medicine}

When the arrangements of NP are obtained by freezing the carrier fluid in colloidal suspensions of NP,\cite{ferrofluids} or by compacting powders of granular solids,\cite{powder} disorder both in dipole orientations and in position result. 
This double disorder  along with the geometric frustration inherent to dipolar interactions may give rise to dipolar spin-glass behavior.
This has been observed experimentally in frozen ferrofluids,\cite{ferrofluids, morup} and in pressed powders of NP.\cite{powder,toro1}

The role played by the orientational disorder (called  texturation) 
has been studied by Monte Carlo simulations in crystaline lattices\cite{russier20} and in random distributions.\cite{alonso19}
In both cases  the  magnetic order at low temperature
changed from FM  to spin-glass (SG) as the orientational disorder increased from textured (parallel axis dipoles) to non--textured (random axis dipoles).
The same has been found in non-textured systems by using the ratio $E_a/E_{dd}$ as disorder parameter.\cite{russier20b}

On the other hand, the role played by the disorder in positions in systems of NP is not completely understood.
Heisenberg dipoles with no local anisotropies placed in random dense packings (RDP) have been studied by Monte Carlo simulations.
Given that the dipoles can rotate freely, the frozen disorder is only in positions and depends on the fraction $\Phi$ of
volume occupied by the NP. Numerical simulations find FM order for $\Phi \gtrsim 0.49$ and SG order otherwise.\cite{alonso20} In the mean-field approach,
Zhang and Widom found FM order for $\Phi \ge 0.295$ under the crude 
approximation of $g(r) = 1$ where $g(r)$ is the radial distribution function.\cite{zhang}
This discrepancy seems to be due to the spatial correlations at short distances for RDP.

Reverting to systems of NP, it must be noticed that the local anisotropy in single domain NP is always non-zero. For this reason, the only way to suppress orientational disorder in such systems in order to explore only positional
disorder effects, is to consider textured systems. It has been recently 
found that the alignment of the easy-axes reduces the dipolar field acting on each NP for volume fractions
$0.3 \lesssim \Phi$.\cite{moya} A certain texturation arises naturally in colloidal liquids even 
in absence of an external field.  Even for moderate values of the volume
  fraction ($ 0.25 \lesssim \Phi \lesssim 0.5 $),  the dipolar hard sphere (DHS) 
fluid polarizes for temperatures below  the ferromagnetic transition temperature 
exhibiting  anisotropic short ranged spatial correlations.\cite{weis, weis2,malherbe23} 

In this paper we investigate the magnetic order of systems of textured dipoles as a function of the positional disorder in RDP.
For this purpose we will use Ising dipoles that point up and down along a common direction such that the structural disorder comes only from
the spatial distribution. For large volume fractions\cite{ayton, alonso19} previous numerical simulations show FM order. Instead, for dipolar
crystals with strong dilution SG order is observed.\cite{PADdilu} We wish to clarify how the degree of spatial disorder in RDP replaces FM order with SG order.

We will study two types of spatially disordered systems of NP.
Firstly we will consider frozen distributions taken from the liquid state of the hard sphere fluid, whose degree of disorder is parametrized
by $\Phi$. Our aim is to obtain the phase diagram in the $T$-$\Phi$ plane and investigate the nature of the low temperature phases. 
The second type of spatially disordered systems studied in this paper are the distributions that arise from freezing  the DHS fluid.
They are obtained by cooling until a   
temperature $T_f$ a fluid of DHS with moderate volume fraction $\Phi$ below  its 
ferromagnetic transition temperature that 
will be denoted by $T_c(DHS,\Phi)$.  Apart from orientational order but
spatial disorder, these spatial distributions show a structural anisotropy that increases when $T_f$
diminishes.\cite{malherbe23} We wish to discover whether such anisotropic configurations favor FM order in textured systems.

Given that we are interested in equilibrium properties, we assume that the dynamics allow Ising dipoles to flip the orientation in order to reach equilibrium. This is tantamount to choosing
a vanishing blocking temperature.\cite{bedanta, fiorani} It is worthwhile to mention that in systems with uniaxial finite anisotropy Monte Carlo simulations indicate the existence of an effective
behavior similar to Ising for $E_a/E_{dd} \gtrsim 30$.\cite{russier20b}

The paper is organized as follows. In Sec.~\ref{mm} we present  the model and  details of the Monte Carlo algorithm, 
and introduce the observables that shall be measured. We present and discuss our results in  Sec.~\ref{results}.
A summary and some concluding remarks are given in Sec.~\ref{conclusion}.

\section{MODEL, METHOD, AND OBSERVABLES}
\label{mm}
\subsection{Model}
\label{models}
We will study systems of $N$ identical NP whose dipoles stay oriented along a common fixed direction $\widehat{a}$. They are labelled with an index $i=1,\dots, N$.
Each NP can be viewed as a sphere of diameter $d$. Its magnetic moment will be denoted by $\vec{\mu}_{i}=\mu s_{i} \widehat{a}$, where $s_i=\pm1$ and $\mu$  takes the
same  value for all spheres. For $s_i=+1$ ($-1$) the dipole points parallel (antiparallel) to $\widehat{a}$.

The $N$ spheres are placed in frozen disordered configurations in a cube of edge $L$. 
The volume fraction occupied by the spheres is $\Phi = N \pi/6 (d^3/L^3)$. We assume periodic boundary conditions.
The position of each particle remains fixed during the simulations and only the signs $s_i$ evolve in time
assuming that the dipoles are able to flip up and down along $\widehat a$.

The Hamiltonian of the system reads
\begin{equation}
{\cal H}= \sum_{ i \ne j}  \varepsilon_d\left( \frac {d}{r_{ij}} \right) ^{3}
\Big( 1 -\frac {3(\widehat{a}\cdot \vec{r}_{ij})( \widehat{a}\cdot \vec{r}_{ij})} {r_{ij}^2}\Big) s_i s_j\;,
\label{eqHamilto}
\end{equation}
where $\varepsilon_d =\mu_{0}\mu^2/(4 \pi d^{3})$ is an energy and $\mu_0$ the magnetic permeability in vacuum.
$\vec{r}_{ij}$ is the position of dipole $j$ as viewed from dipole $i$, and $r_{ij}=\Vert\vec{r}_{ij}\Vert$.
Temperatures will be given in units of $\varepsilon_d/k_{B}$  where $k_{B}$ is the Boltzmann constant.

By the word configuration we will denote a particular realization of positional disorder.  Mathematically it is given by the set
of vectors $\vec{r}_{ij}$ with $i,j=1,\dots,N$, $i\not=j$ and the non-overlapping constraint $r_{ij}>d$ 
 to be plugged in (\ref{eqHamilto}).

We investigate systems of dipoles for two types of configurations. On the one hand we choose configurations of hard spheres corresponding 
to its stable liquid state with given volume fraction $\Phi$. We consider values of  $\Phi$  ranging from diluted systems with $\Phi=0.1$ up to the freezing point ($\Phi=0.49$). 
These configurations are obtained by using the Lubachevsky-Stillinger algorithm,\cite{ls, torquato, donev}
in which the spheres, that are initially very small, are allowed to move and collide while growing in size until they reach the desired value of $\Phi$. 
Their spatial correlations, due to steric effects, are isotropic being 
described by the radial distribution function $g(r)$. 
The amount  of disorder of such configurations is a function of $\Phi$.

The second type of configurations appear by freezing colloidal suspensions of NP.
In practice they are obtained from equilibrium states of the DHS fluid model for low $T_f$ in such a way that the states
correspond to the phase where the system is polarized without crystaline order.\cite{weis, weis2} These configurations
exhibit a large degree of spatial anisotropy which is larger for lower $T_f$.\cite{malherbe23}
The degree of disorder of such configurations is a function of $\Phi$ and $T_f$. 
We have chosen two volume fractions  $\Phi=0.262$ and $\Phi=0.45$  with values of the temperatures $T_f$ adequate to keep the configurations homogeneous.

 Following usual notation in discussions on SG order, we shall call sample 
any system of NP once placed in a specific distribution of fixed positions, that is, in a specific configuration.
Physical results follow by averaging over $N_{s}$ independent samples. These averages are crucial in systems with
strong frozen disorder, for which SG order is expected and where sample-to-sample fluctuations are large. We have used  about $N_s=1000$ samples when FM order was present
and at least $N_s=4000$ samples when SG order was present.  For the simulations reported in Section~\ref{results}.B we have averaged over $N_s=1000$ samples for $\beta_f=0$ and over $250$ for $\beta_f > 0$.

\subsection{Method}

The systems considered here are expected to enter a SG phase in case of strong frozen disorder at low $T$.  As is well-known, SG phases are difficult to simulate
due to the presence of energy landscapes which are beset with barriers and valleys. In order to overcome that difficulty, we used
the tempered Monte Carlo (TMC) algorithm\cite{tempered} which is specially adapted to facilitate  the samples to wander through such rough landscapes in an efficient way.
More specifically, for each sample we run in  parallel $n$ identical
replicas at $n$ different temperatures $T_i$, $i=1,...,n$. After applying $10$ Metropolis sweeps\cite{mc} to each replica, we exchange neighboring pairs of replicas according to detailed balance.
In order to reach equilibrium within a reasonable amount of computer time, we found useful to choose the highest temperature $T_n$ as $T_n \gtrsim 2~T_c$,  and lowest one
$T_1$ as $T_1 \gtrsim0.5 T_c$, where $T_c$ is the expected transition temperature from the PM to the ordered phase. We used an arithmetic distribution of
temperatures
\begin{equation}
  T_i= T_1 + (i-1)\Delta T
\end{equation}
with $\Delta T=0.05$ and  $n\approx 60$ replicas.  When necessary, we added some additional
temperatures with spacing $\Delta T=0.025$ for the low-temperature region $0.5 T_c \lesssim T \lesssim 1.1 T_c$. Only for the systems which are harder to equilibrate
(this occurs with $N=1728$ and $\Phi \le 0.18$) we used the inverse linear distribution\cite{katz19}
\begin{equation}
  \frac{1}{T_i}=\frac{1}{T_1}+\Big(\frac{1}{T_n}-\frac{1}{T_1}\Big)\frac{i-1}{n-1}
\end{equation}
by choosing $n=70$ and $T_1 \approx 0.8~T_c$. The thermal equilibration times $t_{0}$ are estimated following the procedure described in 
Ref.\cite{jpcm17}  For the above list of temperatures we used $t_0=10^6$ Metropolis sweeps for equilibration and took thermal averages for each given
sample within the interval $[t_0,2 t_0]$. A second average over $N_{s}$ samples is needed to obtain physical results. For an observable $u$, this double average will be denoted by $\langle u \rangle$.

The lattice volumes were cubes of edge $L$ with periodic conditions at the boundaries.
The long-range nature of the dipolar-dipolar interaction is treated using Ewald's sums.\cite{ewald} 
Details on the  use of Ewald's sums for dipolar systems are given in Ref.\cite{holm}
We chose $\alpha=4/L$ as  splitting parameter,
computed the sum in real space with a cutoff $r_{c}=L/2$ and the sum in the reciprocal 
space with a cutoff $k_{c}=10$.\cite{holm} 
Given that in case of weak frozen disorder the systems are expected to show FM order,
we performed the Ewald sums by the so-called conducting external conditions, with surrounding permeability $\mu^\prime=\infty$.
This technique allows to avoid shape dependent depolarizing effects.\cite{weis,allen}

\subsection{Observables}
\label{observables}

The main goal of this work has been the determination of the magnetic order as a function of the degree of disorder in the positions of the particles. On general grounds it is
expected that any increase in disorder leads to a reduction of the area occupied by FM order in the phase diagram and an equivalent increase of the area corresponding to SG order.
To characterize both types of magnetic order we employed several observables. First the specific heat
\begin{equation} 
c \equiv \frac{1}{N T^{2}} [\langle{\cal H}^2 \rangle> - \langle{\cal H}\rangle^2 ],
\label{cesp}
\end{equation}
obtained from fluctuations of energy the energy
\begin{equation}
  e\equiv\langle {\cal H}\rangle/N
\end{equation}
  Also, to distinguish FM order, we used the spontaneous magnetization
\begin{equation} 
{m} \equiv \frac{1}{N} \sum_i  s_i \;,
\label{m}
\end{equation}
and evaluated its momenta, $m_{p}=\langle |m|^{p}\rangle$ for $p=1,2,4$ and from them the magnetic susceptibility
\begin{equation}
\chi_{m}\equiv {N\over{k_{B}T}}(m_{2}-m_{1}^{2}),
\label{suscZ}
\end{equation}
and the Binder cumulant
\begin{equation}
B_{m}\equiv{1\over 2} (3-{m_{4}  \over m_{2}^{2} }),
\label{Bm}
\end{equation}
which permits to extract the transition temperature between the phase FM and paramagnetic (PM) phases.

To mark the onset of the SG phase, we evaluated the Edwards-Anderson overlap parameter\cite{ea} defined as
\begin{equation} 
q \equiv \frac{1}{N} \sum_i  s^{(1)}_i s^{(2)}_i,
\label{q}
\end{equation}
for any given sample, where $s^{(1)}_i$ y $s^{(2)}_i$ are the signs of $s_i$ at the site $i$ of two replicas labelled $(1)$ and $(2)$ 
of such sample. Each replica is let to evolve independently at the same temperature. Like for $m$, we also measured the momenta
$q_{p}\equiv \langle |q|^{p}\rangle$ for  $p=1,2,4$ and from them the Binder cumulant
\begin{equation}
B_q\equiv{1\over 2} (3-{q_{4}  \over q_{2}^{2} }).
\label{Bqumulant}
\end{equation}
To identify the transition temperature between PM and SG phases we employed the so-called SG correlation length $\xi_L$ which is given by
\begin{equation} 
\xi^2_L\equiv\frac {1 } {4 \sin^2  ( k /2)}  { \left( \frac{\langle q^2 \rangle} { \langle\mid q(\vec{k})  \mid ^2   \rangle}  -1 \right) }\;, 
\label{phi1}
\end{equation}
where $q(\vec{k})$ is
\begin{equation} 
q({\vec{k}})\equiv \frac{1}{N} \sum_j s^{(1)}_j s^{(2)}_j e^{{\rm i} \vec{k}\cdot \vec{r}_j}\;,
\label{phi2}
\end{equation}
with $\vec{r}_j$ the position of the $j$-th NP, $\vec{k} =(2\pi/L,0,0)$ and $k=\Vert\vec{k}\Vert=2\pi/L$.\cite{longi}. 

Errors in the measurements of these quantities have been calculated as the mean squared deviations of the sample-to-sample fluctuations.

\section{RESULTS}
\label{results}

\subsection{Phase diagram for isotropic HS-like configurations}
\label{ISOphase}

In this section we investigate the magnetic order as a function of the volume fraction $\Phi$ for frozen configurations obtained from equilibrium
states of hard sphere fluids in the range $0 < \Phi \lesssim 0.49$. $\Phi$ measures the degree of spatial disorder on such configurations. We will
show that for decreasing $\Phi$ (which means increasing disorder)  SG order 
replaces the FM order.

\begin{figure}[!t]
\begin{center}
\includegraphics*[width=85mm]{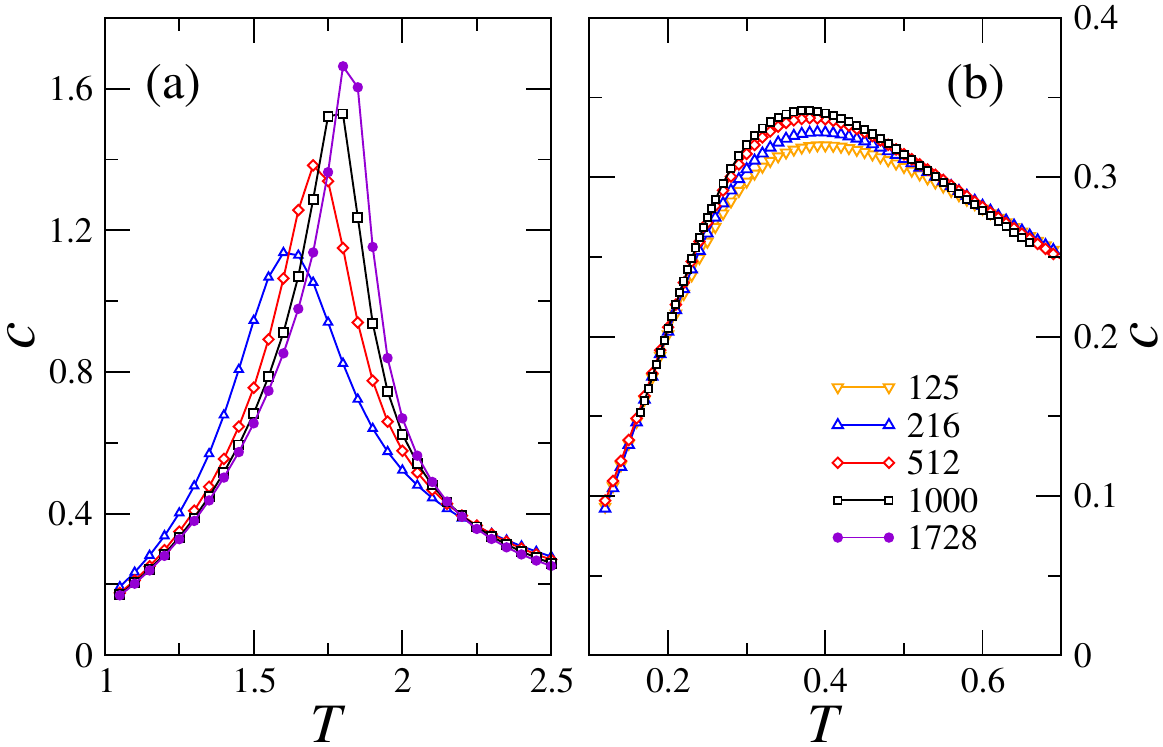}
\caption{(a) Plots of the specific heat $c$ versus $T$ for volume fraction $\Phi=0.4$. Symbols $\smalltriangledown$, $\smalltriangleup$, $\smalldiamond$,
$\smallsquare$, and $\smallblackcircle$ stand for $N=125, 216, 512, 1000$ and $1728$ respectively.  (b) Same as in (a) for volume fraction $\Phi=0.1$.}
\label{figure1}
\end{center}
\end{figure}

\begin{figure}[!b]
\begin{center}
\includegraphics*[width=85mm]{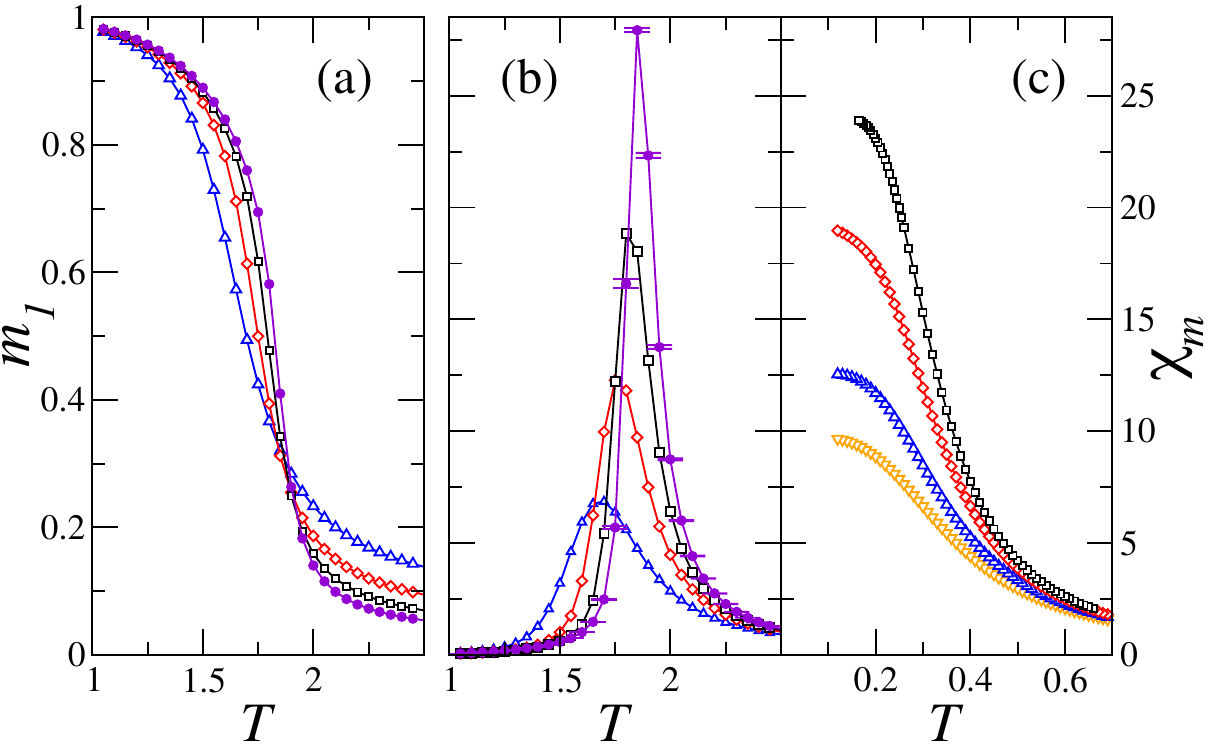}
\caption{(a) Plots of the magnetization $m_1$ versus $T$ for volume fraction $\Phi=0.4$. Symbols $\smalltriangledown$, $\smalltriangleup$, $\smalldiamond$,
$\smallsquare$, and $\smallblackcircle$ stand for $N=125, 216, 512, 1000$ and $1728$ respectively.  
  {\blue b)} Plots of the magnetic susceptibility  $\chi_m$ versus $T$ for volume fraction $\Phi=0.4$. Same symbols as in (a).
  ({\blue c}) Same as in (b) but for volume fraction $\Phi=0.1$.}
\label{figure2}
\end{center}
\end{figure}

A first overview can be grasped from Figs.~\ref{figure1}-\ref{figure2}. Fig.~\ref{figure1}(a) displays plots of the specific heat $c$ vs $T$ for $\Phi=0.4$.
The curves exhibit a marked lambda-shaped peak. Their evident dependence on the number of NP indicates the presence of a singular point in the curve
that corresponds to $N\to \infty$ at $T_c\approx1.9$. That singular behavior is expected in PM-FM second order transitions. Data are consistent with a
logarithmic divergence of $c$ with $N$. Fig.~\ref{figure1}(b) shows the plots obtained for $\Phi=0.1$. In contrast to the previous ones, these plots are
smooth and depend little on the sample size. So, there is no sign of any singular behavior. This is expected in PM-SG transitions with strong structural disorder.

FM order entails the presence of non-vanishing magnetization $m$. Fig.~\ref{figure2}(a) displays $m_1$ vs $T$ for $\Phi=0.4$ at several $N$. They show that
$m_1$ tends to non-zero values for $N \to \infty$ and low $T$, revealing the existence of strong FM order. The curves plotted in Fig.~\ref{figure2}(b) for the
magnetic susceptibility $\chi_m$ vs $T$ confirm this conclusion as they show peaks that become sharper for large $N$. An extrapolation of the positions of
the maxima of those peaks vs $1/N$  provides a value for the transition temperature, $T_c(\Phi=0.4) \simeq 1.9(1)$, in agreement with the estimated $T_c$ obtained from the analysis of Fig.~\ref{figure1}(a). For $T<T_c$ we find that $\chi_m$ does
not diverge with $N$, a fact that validates the above conclusions on FM order.  All that is in contrast to the results obtained for 
$\Phi=0.1$, shown in 
Fig.~\ref{figure2}(c) 
where we see  how the values of $\chi_m$ increase with $N$ for low $T$. Data are consistent with a trend $\chi_m \sim N^p$
for $p\approx 0.45$ and $T \lesssim 0.2$. This behavior suggests
the existence of a SG phase.

Let us discuss now the threshold value of $\Phi$ at which the FM order disappears. Mean-field calculations predict  that FM order persists for $ \Phi \ge \Phi_c = \pi/20 \sim 0.157$.\cite{zhang}

\begin{figure}[!t]
\begin{center}
\includegraphics*[width=82mm]{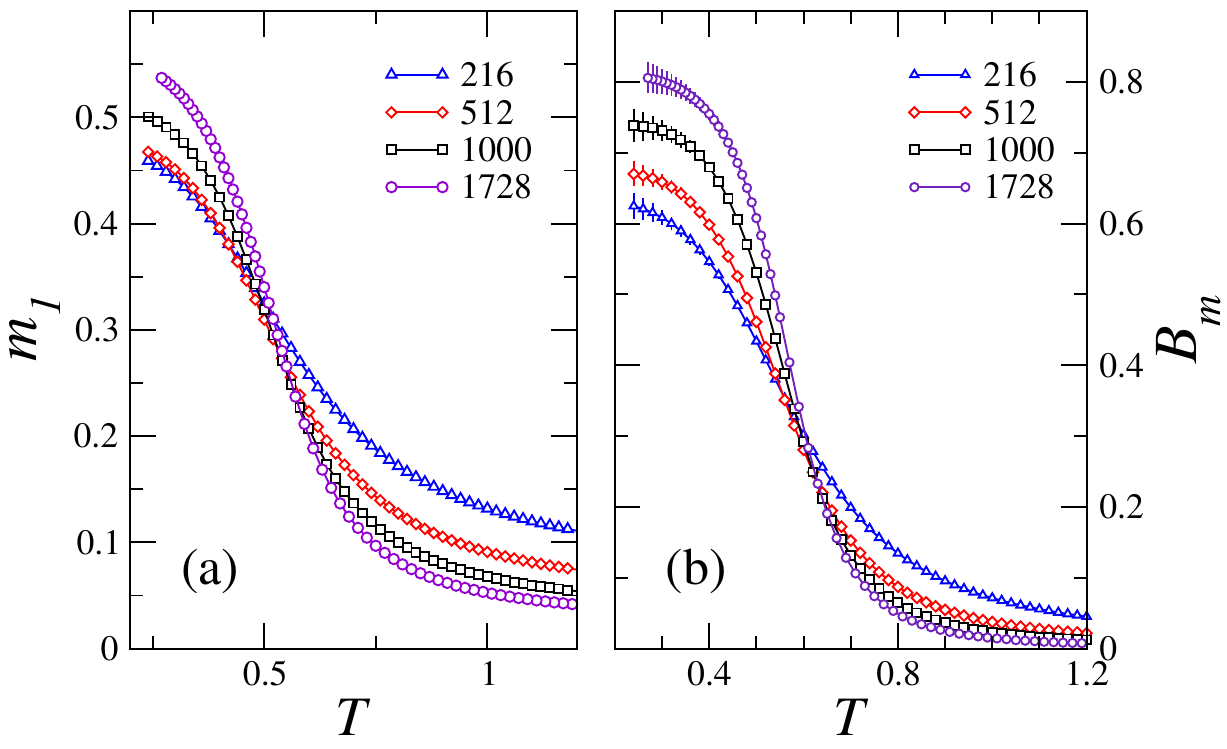}
\caption{(a) Plots of the magnetization $m_1$ vs $T$ for $\Phi=0.18$. Symbols  $\smalltriangleup$, $\smalldiamond$,
$\smallsquare$, and $\smallcircle$ stand for $N=216, 512,1000$ and $1728$ respectively.
(b) Plots of the Binder cumulant of the  magnetization $B_m$ vs $T$ for $\Phi=0.18$. Same symbols as in (a). }
\label{figure3}
\end{center}
\end{figure}

The plots in Fig.~\ref{figure3} show that the FM order  persists at $\Phi=0.18$. The curves of $m_1$ vs $T$ in panel (a) indicate
an  increase in magnetization with $N$ at low $T$, although they also exhibit relevant finite size effects. The Binder parameter of
panel (b) allows to determine the transition temperature within good precision. In general this parameter tends to~1 for $N\rightarrow \infty$ in FM
phases, while from the law of large numbers it follows that in PM phases $B_{m}\rightarrow0$ as $N$ increases. On the other hand, since $B_m$
is dimensionless, it must be independent of $N$ at the critical point. As a consequence, curves of $B_m$ vs $T$ for different values of $N$
cross at $T_c$ for second order transitions. Instead, in presence of an intermediate marginal phase of quasi-long-range FM order, the curves
do not cross but join. Plots of $B_{m}$ vs  $T$ for several $N$ are shown in Fig.~\ref{figure3}(b) for $\Phi=0.18$. Those curves cross at a well
defined critical temperature for $N \ge 512$. \cite{extra} With similar results obtained for $\Phi \ge 0.17$, we can draw a line of transition between PM and FM phases.

\begin{figure}[!t]
\begin{center}
\includegraphics*[width=86mm]{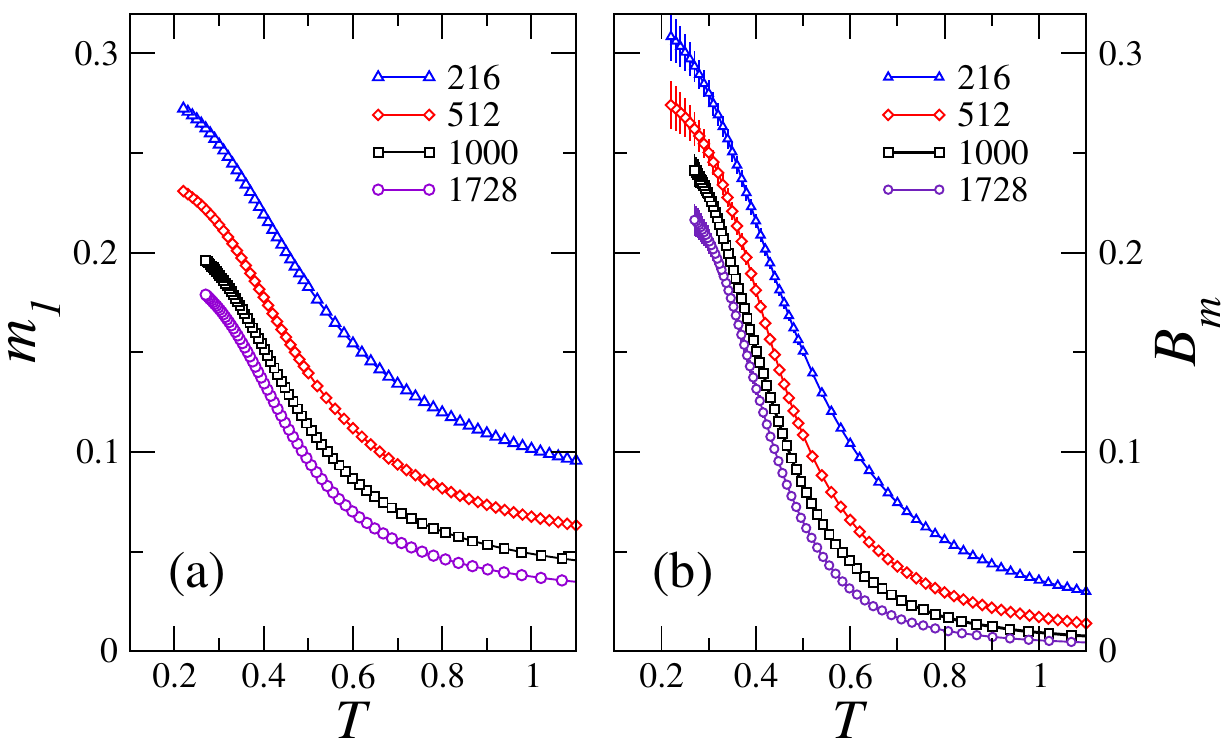}
\caption{(a) Plots of the magnetization $m_1$ vs $T$ for $\Phi=0.14$.
Symbols  $\smalltriangleup$, $\smalldiamond$,
$\smallsquare$, and $\smallcircle$ stand for $N=216, 512,1000$ and $1728$ respectively.
(b) Plots of the Binder cumulant of the  magnetization $B_m$ vs $T$ for $\Phi=0.18$.
Same symbols as in (a).}
\label{figure4}
\end{center}
\end{figure}

The corresponding plots for $\Phi=0.14$ are shown in Fig.~\ref{figure4}. The qualitatively different results illustrate the absence of FM order at
this value of $\Phi$. The plots of the magnetization in panel (a) show that  $m_1$ 
gradually decreases as $N$ increases for all $T$. The data of $m_1$ for low
temperature agree with an algebraic decay $m_1 \sim N^p$ for $p < 1/2$, hence a marginal order is a priori not excluded. However, the plots
of $B_m$ vs $T$ from panel (b) show clearly that $B_m$ vanishes for $N\to \infty$ for all $T$, and this means that the FM order is short
ranged even at low temperature. We have obtained similar plots for all analysed values of $\Phi$ in the range $\Phi \le 0.15$, and this fact excludes FM order.

\begin{figure}[!b]
\begin{center}
\includegraphics*[width=86mm]{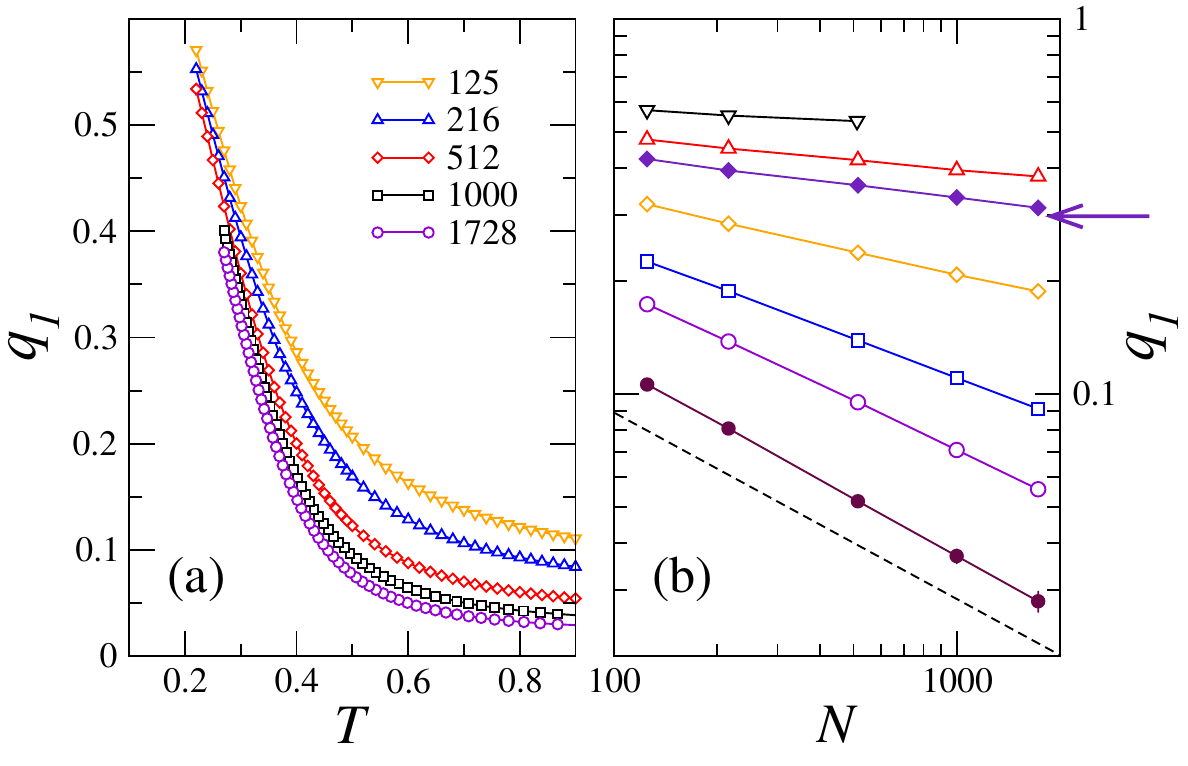}
\caption{  
(a)  Plots of the SG overlap parameter $q_1$  vs  $T$ for $\Phi=0.14$. Symbols $\smalltriangledown$,
$\smalltriangleup$,  $\smalldiamond$, $\square$ and $\smallcircle$  stand for $N=125, 216, 512, 1000$ and $1728$  respectively. 
(b) Log-log plots of $q_1$  vs $N$ for several temperatures at $\Phi=0.14$. From top to bottom, $\smalltriangledown$,
$\smalltriangleup$,  $\smallblackdiamond$, $\smalldiamond$, $\square$ and $\smallcircle$ and $\smallblackcircle$
stand for $T=0.22,~ 0.27,~0.31 ~ 0.37,~ 0.47, ~0.57$ and $0.96$ respectively. 
The arrow marks the data set corresponding to the SG-PM transition temperature.
The dashed line shows the $N^{-1/2}$ decay expected for a paramagnet.}
\label{figure5}
\end{center}
\end{figure}

It is then imperative to  investigate whether at those values of $\Phi$ the FM phase is replaced by SG order. For this purpose we have
evaluated the overlap parameter $q_1$ in Fig.~\ref{figure5}. In panel (a) we show plots of  $q_1$ vs $T$  for $\Phi=0.14$. It is instructive
to compare these plots with those for $m_1$ in Fig.~\ref{figure4}(a) for the same value of $\Phi$. We notice that like for $m_1$, 
also the overlap $q_1$ decreases when $N$ increases for all temperatures.  
To determine if $q_1$ tends to zero in the limit $N\rightarrow \infty$,
we have prepared log-log plots of  $q_1$ vs $N$ in Fig.~\ref{figure5}(b). Data in these plots are consistent with a behavior $q_1\sim 1/N^{p}$
at low temperatures where $p$ is a $T$-dependent exponent. Thus, for example, at $T=0.31$ we have obtained $p\simeq 0.11$. The expected
behavior for a PM phase is $N^{-1/2}$, but we have found it only at high temperature. These properties can be  due to 
the presence
of SG with quasi-long-range order at low $T$. To verify it we have examined the behaviors of $B_q$ and $\xi_L/L$ in terms of $T$. Recall indeed that
in the thermodynamic limit $B_{q} \to 1$ when there is strong SG order, becomes zero in the PM phase, and tends to an intermediate value
at critical points. A similar trend is expected for dimensionless magnitudes like $\xi_L/L$ with a caveat: in case of strong order, this quantity
diverges as $N^{1/2}$ instead of going to~$1$. This makes the splay out of curves for $\xi_L/L$ for different sizes at low temperatures more
prominent than for $B_q$,  and the crossing points are clearer  for second order transitions.\cite{longi}

\begin{figure}[!b]
\begin{center}
\includegraphics*[width=86mm]{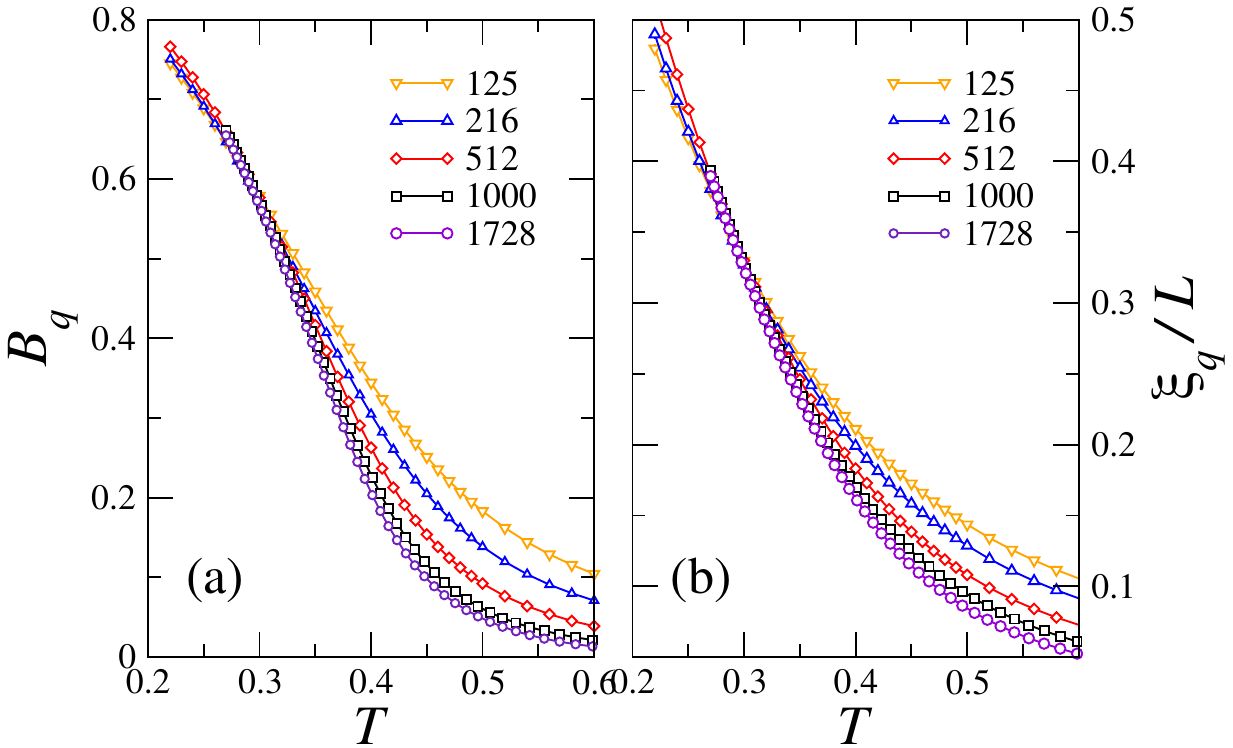}
\caption{(a)  Plots of the SG correlation length $\xi_L/L$  vs  $T$ for $\Phi=0.14$. Symbols $\smalltriangledown$,
$\smalltriangleup$,  $\smalldiamond$, $\square$ and $\smallcircle$  stand for $N= 125, 216, 512, 1000$ and $1728$  respectively. 
(b) Plots of the SG correlation lenght $\Xi_L/L$  vs  $T$ for $\Phi=0.14$. Same symbols as in (a). 
}
\label{figure6}
\end{center}
\end{figure}

The curves $B_{q}$ vs $T$ for $\Phi=0.14$ shown in Fig.~\ref{figure6}(a) merge at $T$ below a certain
value $T_{sg} \simeq 0.31(2)$, rather than crossing. The spread that those curves exhibit for $T < T_{sg}$ becomes almost zero for $N \ge 512$. Thus, $B_q$
does not tend to~1 in the thermodynamic limit. Then, the curves $B_{q}$ collapse for $N \to \infty$ and  $T \le T_{sg}$, and this is
consistent with the algebraic decay found for $q_{1}$. The plots of $\xi_L/L$ vs $T$ in Fig.~\ref{figure6}(b)  exhibit a similar behavior.
All that emphasizes that there exists SG order with quasi-long-range order, as it happens in other systems with NP and strong
structural disorder.

 The temperature $T_{sg}$ that marks the transition between PM and SG orders is a function of $\Phi$ and for that reason it will be represented as $T_{sg}(\Phi)$.
The fact that the merging of different curves be dominant over crossing makes the determination of $T_{sg}(\Phi)$
less precise than for the PM-FM transition. In any case $T_{sg}(\Phi)$ is quite smooth as a function of $\Phi$ for strong dilution.
For $\Phi=0.1$ we have obtained $T_{sg}/\Phi = 1.9(1)$, in agreement with the relation $ T_{sg} = x$ found in the limit of strong dilution
for systems of dipoles in 
crystalline 
simple cubic (SC) lattices with fraction $x$ of occupied sites.\cite{PADdilu} For a diluted system of spheres with SC order, this
relation reads $T_{sg}/\Phi =1/\Phi_{SC}\simeq 1.91$, where $\Phi_{SC}$ is the volume fraction for SC lattices.

\begin{figure}[!t]
\begin{center}
\includegraphics*[width=86mm]{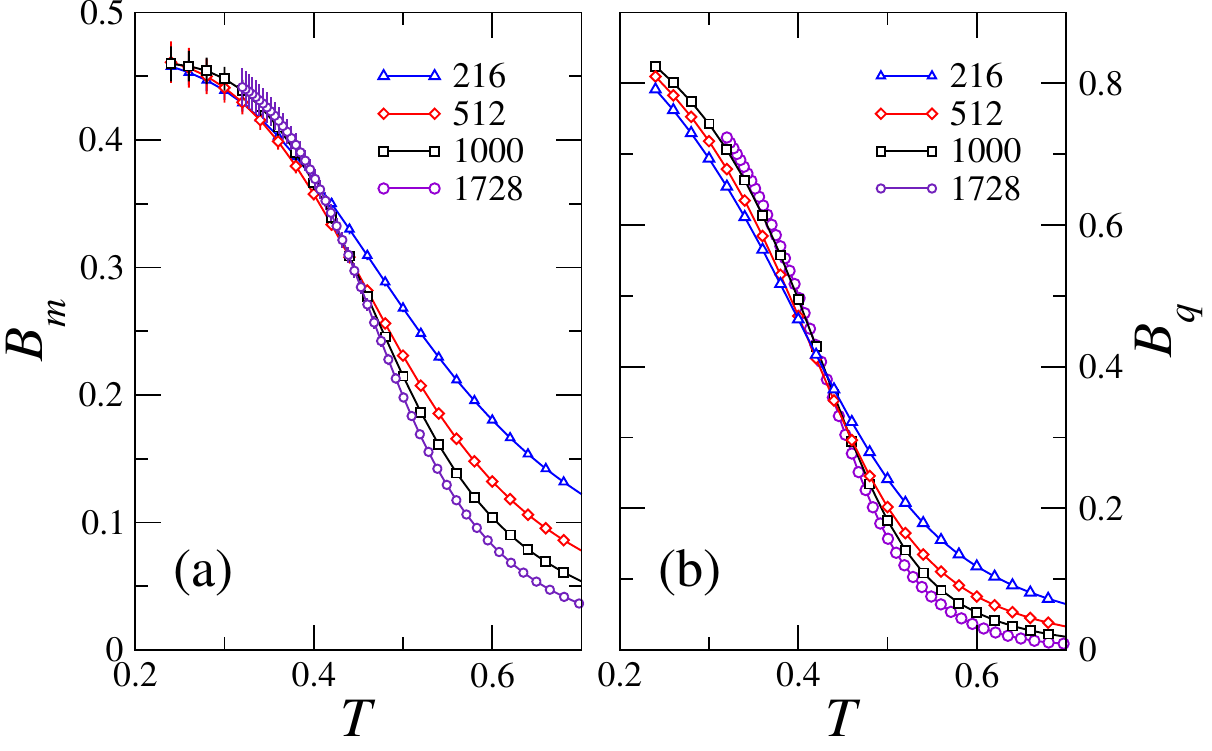}
\caption{(Color online)
  (a) Plots of the Binder cumulant $B_{m}$  vs  $T$ for $\Phi=0.16$. 
  $\smalltriangleup$,  $\smalldiamond$, $\square$ and $\smallcircle$  stand for systems with $N=216, 512, 1000$ and $1728$ dipoles respectively.
  (b) Plots of the Binder cumulant for the overlap parameter $B_{q}$  vs $T$ for $\Phi=0.16$. Same symbols as in (a). 
}
\label{figure7}
\end{center}
\end{figure}

As a last step we determine the low temperature boundary of the FM phase. Mean-field theory predicts the onset of FM order for $T=0$ at $\Phi_c=0.157$.
Let us examine first the data obtained for $\Phi=0.16$. Plots of $B_{q}$ vs $T$ for various sizes are shown in Fig.~\ref{figure7}(b). We estimate 
that the curves cross at $T_c \approx 0.44$. For $T<T_c$ we have found (not shown) that $q$ does not vanish in the thermodynamic limit, a fact that points to
the presence of SG order. Contrarily, the curves of $B_{m}$ vs $T$ shown in panel (a) of the same figure merge at low temperatures. In particular, the curves
for $N=1000$ and $N=1177$ fall on top of each other within the error bars for $T<T_c$. This suggests that the transition line between FM and SG lies at $\Phi \approx 0.16$
at low temperature.

 \begin{figure}[!b]
\begin{center}
\includegraphics*[width=86mm]{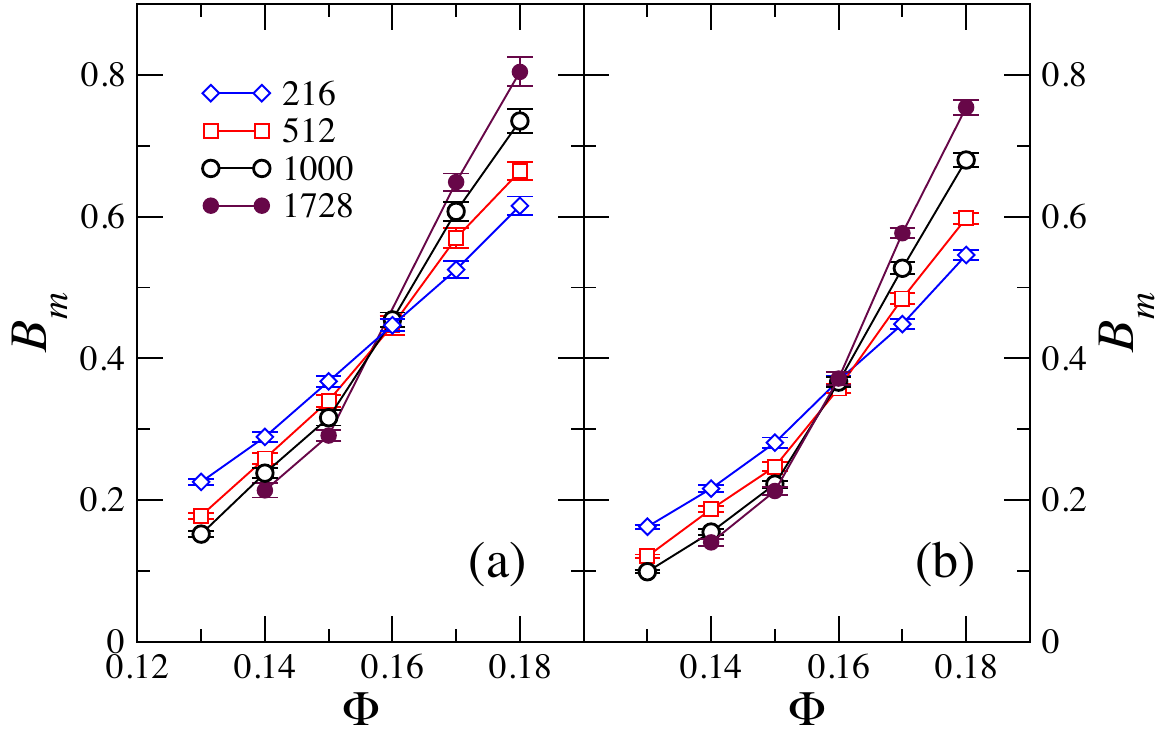}
\caption{   (a) Plots of the Binder cumulant $B_{m}$  vs  $\Phi$ for $T=0.28$. 
$\smalldiamond$, $\smallsquare$ and $\smallcircle$  and $\smallblackcircle$ stand for systems with $N=216, 512, 1000$ and $1728$ dipoles respectively.
 (b)The same as in (a) but for temperature  $T=0.40$.
}
\label{figure8}
\end{center}
\end{figure}

This line of critical points can be recovered in a more precise way with data obtained from simulations in the range $0.13<\Phi< 0.18$.
To this end, we prepare plots of $B_m$ vs $\Phi$ at different $N$ along isotherms for $T$ below the PM region. Fig.~\ref{figure8} shows the plots
for the isotherm at $T= 0.28 $ in panel (a) and $T=0.4$ in panel (b). Recall that $B_m$ diminishes as $N$ increases in the SG phase while in the FM
phase $B_m$ increases with $N$. As shown by both panels in the figure, we have found that the curves of $B_m$ vs $\Phi$ cross at the transition $\Phi_c(T)$.
A very precise result can be obtained if we have many values of $\Phi$ available. 
The transition line $\Phi_c(T)$ is almost vertical at $\Phi = 0.160(5)$ in good agreement with
mean-field approximation. We also notice the  well defined 
separation in the curves above and below the cross point in the plots of Fig.~\ref{figure8}. This detail
rules out the possibility of the presence of critical phases in between FM and SG in the region close to $\Phi \simeq\Phi_c(T)$.

\begin{figure}[!b]
\begin{center}
\includegraphics*[width=82mm]{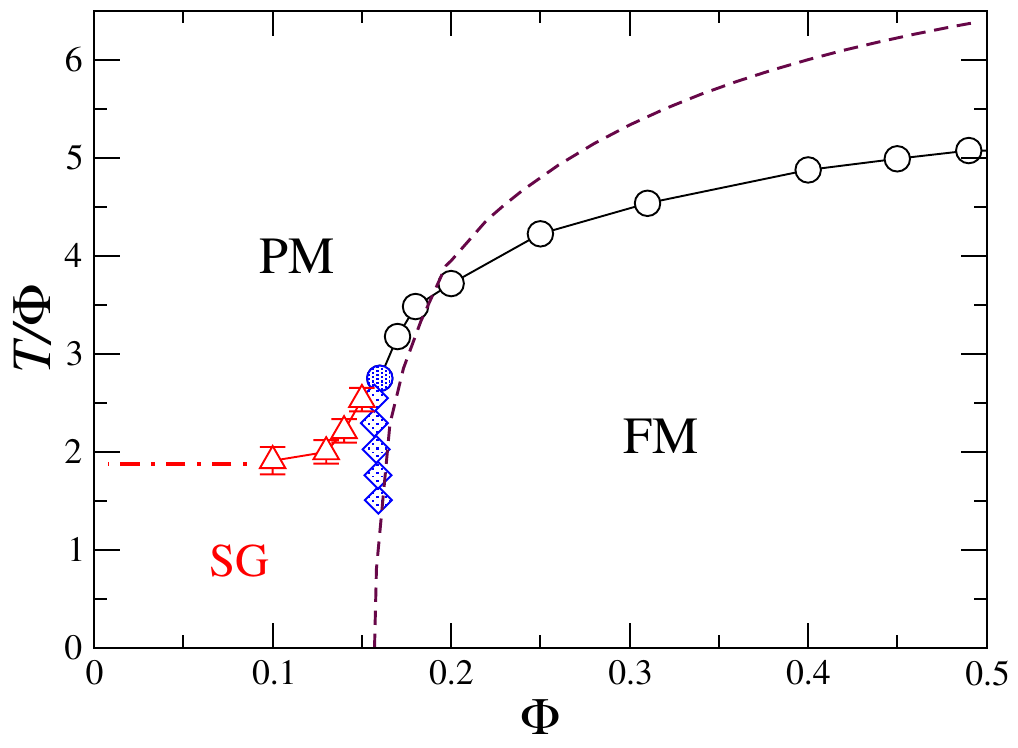}
\caption{Phase diagram in the plane $(T,\Phi)$ for the fully textured dipolar Ising model on random HS-like configurations. Symbols $\medcircle$ stand for PM-FM transition
obtained from the $B_{m}$ vs $T$ plots. Symbols $\triangle$ stand for PM-SG transition
obtained from the $B_{q}$ vs $T$ plots.  Symbols $\meddiamond$ represent the FM-SG transition and follow
from the $B_{m}$ vs $\Phi$ plots. Error bars are smaller than the size of these symbols. The  dashed line stand for a mean-field
calculation by Zhang-Widom.\cite{zhang} The horizontal red dashed line comes from previous calculations for strongly diluted DIS in crystals.}
\label{figure9}
\end{center}
\end{figure}

The results of this section can be gathered in the phase diagram of Fig.~\ref{figure9} which shows the extension of the several FM, SG and PM regions.
They are separated by second order transition lines. The slope of the transition line at low density is fairly zero, in such a way that the ratio $T_{sg}/\Phi$ takes a fixed value, $T_{sg}/\Phi \simeq 1.9$.
Mean-field theory yields a good approximation of the boundary line of the FM phase at low temperatures.\cite{zhang} This approximation is carried out by assuming fully random particle positions
which is in contrast with the results obtained for systems of dipoles with no local anisotropy,\cite{alonso20} for which the onset found at $\Phi=0.49$ coincides with the freezing point
of the hard sphere fluid. This onset depends on the details of the radial distribution function for short distances.\cite{doublehorn} We end this discussion by noticing that differently from systems
of Ising dipoles with orientational disorder,\cite{alonso19}  we have found for the systems studied in this paper no trace of reentrances or other intermediate phases, as clearly shown in Fig.~\ref{figure9}.

\subsection{FM order on anisotropic DHS fluid-like configurations}
\label{FManiso}

In this subsection we will describe the results obtained by exploring the FM order of textured Ising dipoles placed in frozen DHS fluid-like distributions of particles. 
These positional distributions are taken from equilibrium configurations of the DHS fluid at low temperatures $T_f$.\cite{malherbe23} 
We consider volume fractions in the range $ 0.25 \lesssim \Phi \lesssim 0.5 $ for which the DHS fluid is known 
to polarize below the transition temperature, $T_c(DHS,\Phi)$.\cite{weis, weis2}  
Within this  interval of values of $\Phi$  and 
for a wide range of temperatures $T_f$  below $T_c(DHS,\Phi)$ the equilibrium configurations for the DHS fluid  
exhibit some partial alignment of the magnetic moments ${\widehat{\mu}_i}$ along a common 
direction  and a certain degree of anisotropy in the positional distribution.
At the same time these  fluid-like configurations are still homogeneus and show absence of crystalline order.\cite{weis2} 
It is this type of configurations that arise naturally in colloidal suspensions of particles at low temperature.

Any of those equilibrium configurations is determined by the positions $\vec{r}_i$ and the instantaneous
orientations of the magnetic moments $\widehat{\mu}_i$ of all particles, $i=1,\dots,N$. At low $T_f$ 
the orientations exhibit nematic order along a preferred direction $\widehat a$. In this case the configurations
are said to be {\it partially textured}.

As  it is customary when studying nematic order, the direction $\widehat a$  can be determined as the eigenvector related 
to the largest eigenvalue of the so-called nematic tensor \cite{allen}
\begin{equation}
 \bar{Q}_n = \frac{1}{2N} \sum_i (3{\widehat{\mu}_i}{\widehat{\mu}_i} - \bar{I}) \;.
\end{equation}
Thus, the degree of  texturation
can be quantified by the value of the nematic order parameter $\lambda_n$. Similarly, the degree of anisotropy 
in the disordered positional distribution can be assessed by the {\it structural} nematic order parameter  $\lambda_s$ associated with the tensor

  \begin{equation}
  \bar{Q}_s = \frac{1}{2N_{nn}} \sum_{nn} (3{\widehat{r}_{nn}}{\widehat{r}_{nn}} - \bar{I})\;,\
  \label{tomeueq}
  \end{equation}
where $\widehat{r}_{nn}$ are the normalized relative positions $\widehat{r}_{ij}\equiv\vec{r}_{ij}/{r}_{ij}$ between 
pairs of particles     whose $r_{ij}$
distance is smaller than 
a threshold value chosen as $r_s=1.2d$,   
and $N_{nn}$ is the number of such pairs. \cite{holdsworth}
$\lambda_s$ is the largest eigenvalue of $ \bar{Q}_s$
and measures the degree of aligment of the set of $\widehat{r}_{nn}$ along a preferred direction.

\begin{figure}[!t]
\begin{center}
\includegraphics*[width=84mm]{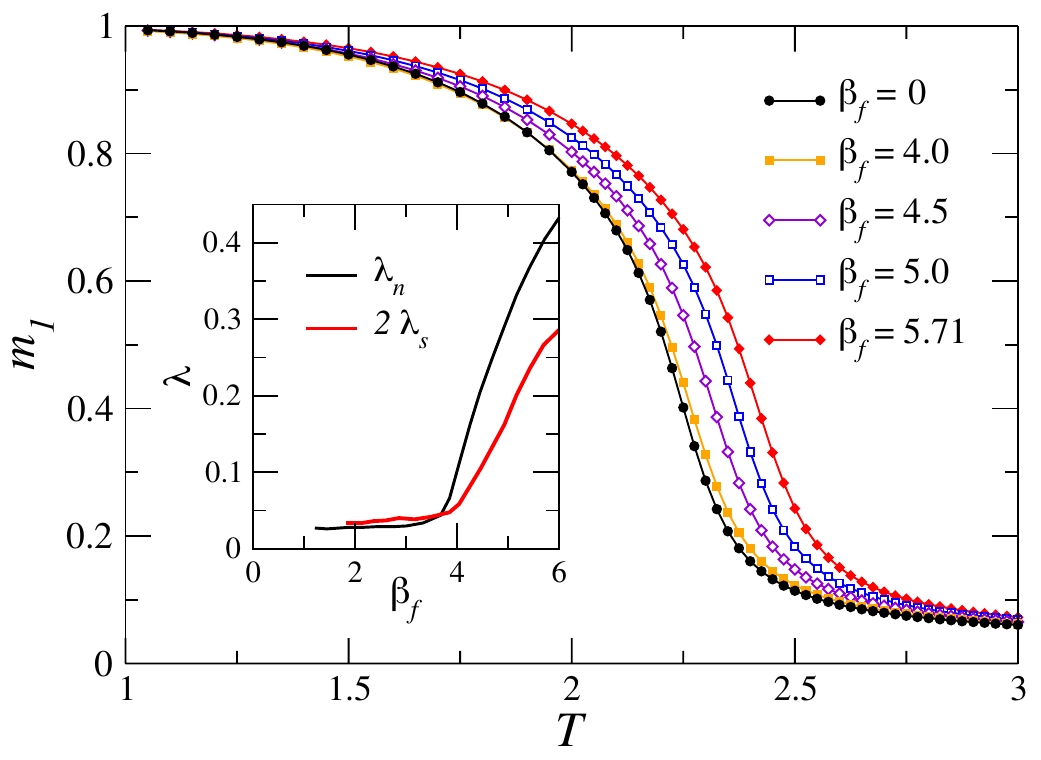}
\caption{  Plots of  the magnetization  $m_1$  vs   $T$ for ensembles of $N=1177$ dipoles placed on anisotropic frozen configurations obtained for $\Phi=0.45$ and the inverse freezing temperatures $\beta_f$ indicated in the figure.
The inset shows the structural $\lambda_s$ and the orientational $\lambda_n$ 
nematic order parameter for the DHS fluid versus the inverse temperature $\beta_f$ for density  $\Phi=0.45$ and $N=1177$ (data taken from Ref.~\onlinecite{malherbe23}). 
}
\label{figure10}
\end{center}
\end{figure} 

\begin{figure}[!b]
\begin{center}
\includegraphics*[width=84mm]{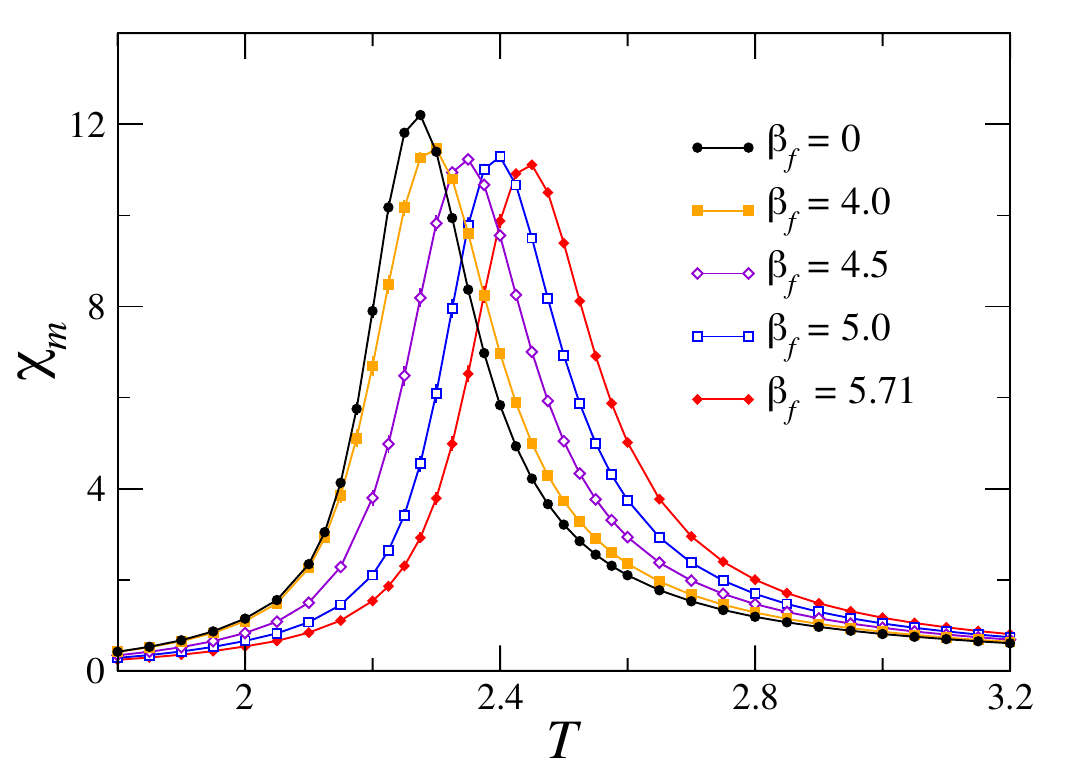}
\caption{ Plots of  the susceptibility  $\chi_m$  vs   $T$ for ensembles of $N=1177$ 
dipoles placed on anisotropic frozen configurations obtained for $\Phi=0.45$ and the inverse freezing temperatures $\beta_f$
 indicated in the figure.
}
\label{figure11}
\end{center}
\end{figure}
 
The behavior of those eigenvalues has been explored for  $\Phi=0.45$ and $\Phi=0.262$ in Ref.~\onlinecite{malherbe23}.
Plots of  $\lambda_n$ and $\lambda_s$ versus the inverse temperature $\beta_f=1/T_f$  are given for $\Phi=0.45$ and $N=1177$
in the inset of Fig.~\ref{figure10}. The plateaux found for $\beta_f  \lesssim \beta_c(DHS,\Phi) = 4$ with
$\lambda_s \simeq \lambda_n \simeq 0$ indicate that the configurations remain isotropic for temperatures above the PM-FM transition,
(the notation $\beta_c(DHS,\Phi)\equiv1/T_c(DHS,\Phi)$ has been used). In contrast, for  
$\beta_f \gtrsim 4$ both $\lambda_n$ and $\lambda_s$ increase with $\beta_f$, indicating that the double anisotropy strengthens as temperature is lowered. For
$\Phi=0.262$, where  $\beta_c(DHS,\Phi) = 7.7$, the behavior is qualitatively the same, apart from the fact that
there seems to form chains of spheres at very low temperature instead of the homogeneous configurations observed for $\Phi=0.45$.

\begin{figure}[!b]
\begin{center}
\includegraphics*[width=84mm]{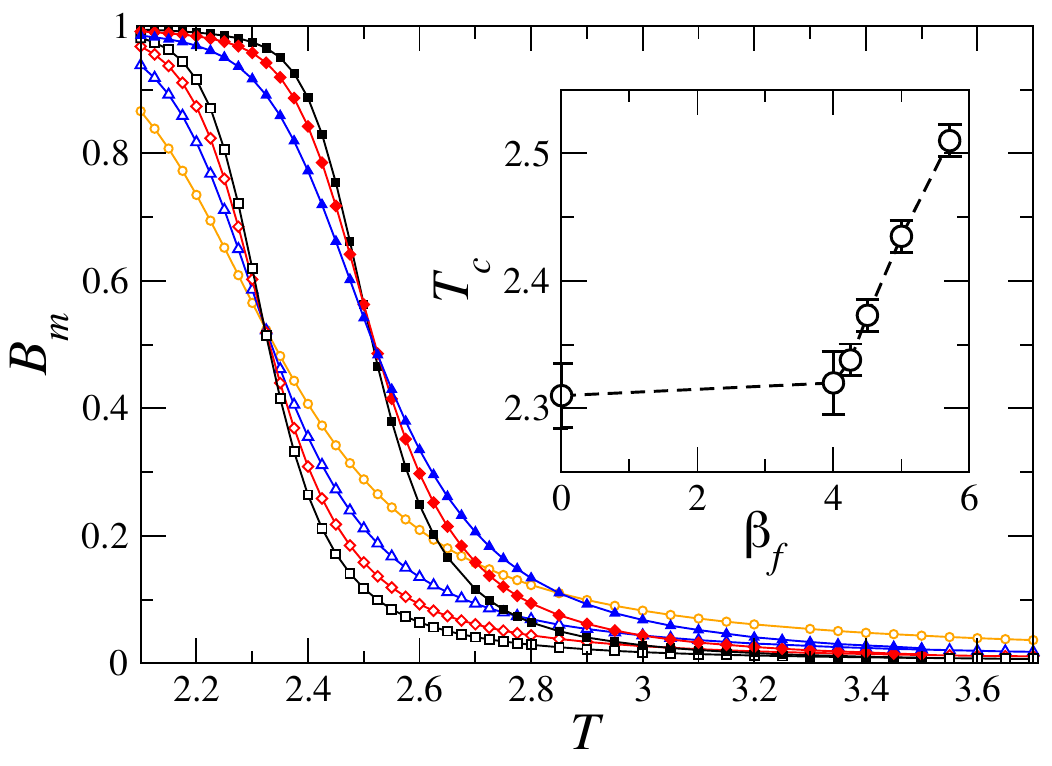}
\caption{ (a) Plots of  $B_{m}$  vs  $T$ for systems of $N$ dipoles placed on anisotropic frozen configurations obtained for
$\Phi=0.45$ and given freezing temperatures $\beta_f$.   
$\smallblacktriangleup$, 
$\smallblackdiamond$ and  $\smallblacksquare$  
stand for $\beta_f=5.71$ and system sizes
$N=453, 758$ and $1177$ respectively. 
$\smallcircle $, $\smalltriangleup$, 
$\smalldiamond$ and  $\smallsquare$ stand for $\beta_f=0$ and 
$N=125, 216, 512$ and $1000$ respectively. 
The solid lines are guides to the eye.
The inset shows the PM-FM transition temperature $T_c$ 
vs $\beta_f$ for volume fraction $\Phi=0.45$.
}
\label{figure12}
\end{center}
\end{figure}

Up to here, we have described equilibrium DHS fluid-like configurations at low $T$ which exhibit partial texturation 
in the orientations of the dipoles as well as structural anisotropy in their positions.\cite{malherbe23}
 However, given that this work is aimed at the study of the role of frozen positional disorder, we proceed by studying fully textured systems of DHS fluid-like configurations in equilibrium.
Once picked one of these configurations, i.e., once the sets of positions $\vec{r}_i$ and momenta $\widehat{\mu}_i$ are 
fixed throughout the lattice, the frozen textured distribution is built
 by firstly computing the nematic tensor $\bar{Q}_n$ and secondly by choosing the nematic 
 director $\widehat a$ of $\bar{Q}_n$ as the common direction along which all the Ising dipoles 
  are placed.

The question now is how the remaining  structural positional anisotropy in those systems affects the order at low $T$. We do not expect
to find any SG order for the volume fractions considered here ($ 0.25 \lesssim \Phi \lesssim 0.5 $). Note from the previous section, that for
$\Phi \lesssim 0.16$ only FM order is expected at low $T$ even for isotropic configurations.  Thus, by studying DHS
systems we intend to analyze whether the structural positional anisotropy enhances the FM order already present in isotropic HS configurations.

Curves of magnetization $m_1$ vs $T$ are shown in Fig.~\ref{figure10} for $\Phi=0.45$ and $N=1177$ at various
values of $\beta_f =1/T_f$. The system 
at low temperature
is in a FM phase even for $\beta_f=0$ in absence of anisotropy, in agreement with the previous subsection.
The result for $\beta_f \simeq \beta_c(DHS,\Phi) \simeq 4$ is practically the same that for $\beta_f=0$. Only for $\beta_f > 4$ we see
the curves of magnetization to move rightwards as $\beta_f$ increases, a fact that indicates that the increase in anisotropy favors FM order.
The susceptibility $\chi_m$ vs $T$ curves shown in Fig.~\ref{figure11}  exhibit peaks which
are typical for ferromagnets. The positions of those peaks for $\beta_f > 4$ shift to the right as $\beta_f$ is increased,
indicating that the  PM-FM  transition temperature 
$T_c$ increases with the anisotropy.

A precise determination
of $T_c$ can be obtained from the crossing points of the Binder parameter vs $T$ for different
sizes, as shown in Fig.~\ref{figure12}. 
The inset in this figure  shows the transition temperature $T_c$ vs $\beta_f$ for $\Phi=0.45$.
This figure is to be compared with the inset of Fig.~\ref{figure10}. We can appreciate that $T_c$ increases with $\beta_f$
only for $\beta_f > \beta_c(DHS,\Phi) \simeq 4$ when the structural anisotropy quantified by $\lambda_s$ increases. It is worth
to mention that for $\beta_f \gtrsim 4$ the  values of  $T_c$ obtained here for fully textured systems along the nematic director $\widehat a$
practically coincide with the values of  $T_c$  found for ensembles of Ising dipoles placed on the same frozen
configurations $\{\vec{r}_i\}$ but pointing up or down along the
orientations $\{\widehat{\mu}_i\}$ of the original DHS configurations.
\cite{malherbe23}  This suggests that the
procedure used in this work for imposing a common direction $\widehat a$ (obtained from the set of values $\widehat{\mu}_i$) to build fully textured samples
keeps all the relevant information for the FM order induced by the structural anisotropy on positions.
In other words, the fluctuations of the Ising axes around the mean value $\widehat{a}$ play only a marginal role.

\begin{figure}[!t]
\begin{center}
\includegraphics*[width=86mm]{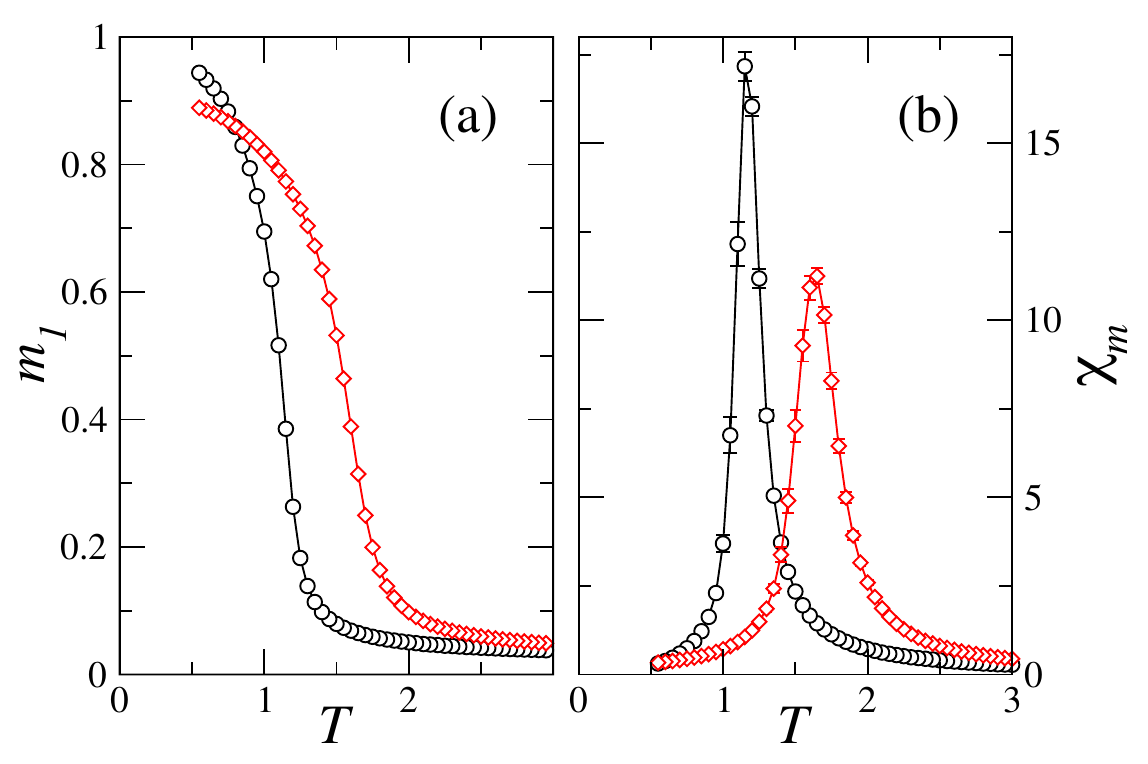}
\caption{  (a) Plots of  the magnetization  $m_1$  vs   $T$ for ensembles
of $N=1000$  dipoles placed on anisotropic frozen
configurations obtained for $\Phi=0.262$.  Symbols $\smallcircle$ and
$\smalldiamond $  stand for $\beta_f=0$ and $8.5$ respectively.  (b)
Plots of  the susceptibility  $\chi_m$  vs   $T$ for ensembles of
$N=1000$,  for $\Phi=0.262$. Same symbols as in (a).}
\label{figure13}
\end{center}
\end{figure}

For the volume fraction $\Phi=0.262$ similar results are obtained. Fig.~\ref{figure13} exhibits curves of
$m_1$ and $\chi_m$ vs $T$ for $\beta_f=0$ and $\beta_f=8.5$  (a value larger than $\beta_c(DHS,\Phi)=7.7$).
Both $m_1$ and $\chi_m$ curves move to the right as $\beta_f$ is increased, indicating again that the presence of  
structural  anisotropy favors FM order.
However it can be noticed that for very low temperatures the magnetization $m_1$ for $\beta_f=8.5$ is lower than for 
$\beta_f=0$. This fact can be related to the formation
of inhomogenities in the DHS fluid for low $T_f$.\cite{weis2, malherbe23}

\section{CONCLUSIONS }
\label{conclusion}

We have studied by Monte Carlo simulations the effect of positional disorder on the collective properties
of fully textured systems of identical magnetic nanospheres that behave as Ising dipoles along common easy axes.

We have first studied frozen isotropic systems of hard spheres obtained along the stable liquid branch with volume fraction $\Phi$
ranging from low values up to the freezing point ($\Phi \simeq 0.49$). By analysing the phase diagram on the $T$-$\Phi$ plane, we have
found a low-temperature ferromagnetic phase for $\Phi \gtrsim \Phi_o= 0.160(5)$ in good agreement with mean-field calculations that
assume complete randomness in positions. This phase exhibits strong long-range order. For $\Phi \less \Phi_o$ this ferromagnetic phase
disappears giving rise to a spin-glass phase for temperatures below $T_{sg}(\Phi)$. For strong dilution we find $T_{sg}(\Phi)/\Phi =  1.9(1)$.
The nature of the  dipolar spin-glass phase is similar to the one observed in other systems of Ising dipoles with strong frozen disorder.
Plots of Binder cumulants vs $\Phi$ allow to obtain the transition line between the ferromagnetic and spin-glass phases. We find neither
an appreciable reentrance nor an intermediate region with quasi-long-range ferromagnetic between the FM and SG phases.  

We have also studied anisotropic spatial systems for $\Phi=0.45$ and $\Phi=0.262$. They have been obtained by freezing 
the liquid state of the dipolar hard sphere fluid in its polarized state at sufficiently low temperatures $T_f$. Such systems
develop some texturation as well as anisotropic spatial correlations that increase as $T_f$ is decreased.  The ferromagnetic
order of parallel dipoles placed on such configurations along their nematic director is enhanced as $T_f$ decreases.

\section*{Acknowledgements}

We thank the Centro de Supercomputaci\'on y Bioinform\'atica  at University of M\'alaga,  and the
Institute Carlos I at University of Granada for their generous allocations of computer time.  
J.J.A. also thanks the Italian ``Fondo FAI'' for financial support and the warm hospitality received during his stay in the Pisa INFN section.


\begin{thebibliography}{99}
\bibitem{extra} For some values of $\Phi$, pairs of curves do not cross precisely at the same
point, but $1/L$ extrapolations of the crossing points allow to obtain well defined values of $T_c$.
\bibitem{doublehorn} For volume fractions below $\Phi\approx0.49$ the HS fluid is in a stable liquid phase.
Compressing rapidly this fluid results in the appearance of metastable branches of amorphous states with densities
up to  $\Phi \approx 0.64$. For this glassy states with $\Phi \ge 0.49$, the radial distribution function 
$g(r)$ exhibits characteristic double-horn peaks at $r=\sqrt{3}d$ and $2d$ where $d$ is the diameter or size of the NP. In absence of local anisotropies,
Monte Carlo simulations find that the existence of FM for systems of dipoles in RDP depends on the presence of this double-horn.
\end{thebibliography}
\end{document}